\documentclass[sigconf]{acmart}

\AtBeginDocument{%
  \providecommand\BibTeX{{%
    \normalfont B\kern-0.5em{\scshape i\kern-0.25em b}\kern-0.8em\TeX}}}

\copyrightyear{2022} 
\acmYear{2022} 
\setcopyright{acmcopyright}\acmConference[WWW '22]{Proceedings of the ACM Web Conference 2022}{April 25--29, 2022}{Virtual Event, Lyon, France}
\acmBooktitle{Proceedings of the ACM Web Conference 2022 (WWW '22), April 25--29, 2022, Virtual Event, Lyon, France}
\acmPrice{15.00}
\acmDOI{10.1145/3485447.3512094}
\acmISBN{978-1-4503-9096-5/22/04}

\usepackage{multirow}
\usepackage{footmisc}
\usepackage{subcaption,siunitx,booktabs}
\usepackage{caption}
\usepackage{bbm}
\usepackage{todonotes}
\usepackage{amsmath}

\usepackage{balance}
\usepackage{enumitem}
\usepackage[ruled]{algorithm2e}
\SetKwComment{Comment}{$\triangleright$\ }{}

\SetKwInput{KwInput}{Input}                
\SetKwInput{KwOutput}{Output}              

\renewcommand{\todo}[1]{\iffalse #1 \fi[TODO]}

\begin{document}

\title{Re4: Learning to Re-contrast, Re-attend, Re-construct for Multi-interest Recommendation}


\author[S. Zhang*, L. Yang*, D. Yao*, Y. Lu, F. Feng, Z. Zhao, T. Chua,  F. Wu]{
    Shengyu Zhang$^{1*}$, Lingxiao Yang$^{1*}$, Dong Yao$^{1*}$, Yujie Lu$^{6}$, Fuli Feng$^{4}$, Zhou Zhao$^{1,2\dagger}$, \\ Tat-seng Chua$^{5}$, Fei Wu$^{2,3\dagger}$
}
\affiliation{
    $^1$ Zhejiang University \ $^2$ Shanghai Institute for Advanced Study of Zhejiang University \country{}
}
\affiliation{
	$^3$ Shanghai AI Laboratory \ $^4$ University of Science and Technology of China\country{}
}
\affiliation{
	$^5$ National University of Singapore \ $^6$ University of California, Santa Barbara, United States\country{}
}

\email{
  {sy_zhang, yaodongai, zhaozhou, wufei}@zju.edu.cn
}
\email{
	{yujielu10,fulifeng93,shawnyang1110}@gmail.com
}
\email{dcscts@nus.edu.sg}
\renewcommand{\authors}{Shengyu Zhang, Lingxiao Yang, Dong Yao, Yujie Lu, Fuli Feng, Zhou Zhao, Tat-seng Chua, Fei Wu}

\newcommand{\etal}{\textit{et al}.}
\newcommand{\ie}{\textit{i}.\textit{e}.}
\newcommand{\eg}{\textit{e}.\textit{g}.}
\newcommand{\wrt}{\textit{w}.\textit{r}.\textit{t}. }
\newcommand{\vpara}[1]{\vspace{0.05in}\noindent\textbf{#1 }}
\newcommand\norm[1]{\left\lVert#1\right\rVert}

\renewcommand{\shortauthors}{Shengyu Zhang, Lingxiao Yang, Dong Yao, Yujie Lu, Fuli Feng, Zhou Zhao, Tat-seng Chua, \& Fei Wu}
\renewcommand{\thefootnote}{\fnsymbol{footnote}}

\begin{abstract}
Effectively representing users lie at the core of modern recommender systems. Since users' interests naturally exhibit multiple aspects, it is of increasing interest to develop multi-interest frameworks for recommendation, rather than represent each user with an overall embedding. Despite their effectiveness, existing methods solely exploit the encoder (the forward flow) to represent multiple aspects of interests. However, without explicit regularization, the interest embeddings may not be distinct from each other nor semantically reflect representative historical items. Towards this end, we propose the Re4\footnote{Code available at \url{https://github.com/DeerSheep0314/Re4-Learning-to-Re-contrast-Re-attend-Re-construct-for-Multi-interest-Recommendation}} framework, which leverages the backward flow to reexamine each interest embedding. Specifically, Re4 encapsulates three backward flows, \ie, 1) Re-contrast, which drives each interest embedding to be distinct from other interests using contrastive learning; 2) Re-attend, which ensures the interest-item correlation estimation in the forward flow to be consistent with the criterion used in final recommendation; and 3) Re-construct, which ensures that each interest embedding can semantically reflect the information of representative items that relate to the corresponding interest. We demonstrate the novel forward-backward multi-interest paradigm on ComiRec, and perform extensive experiments on three real-world datasets. Empirical studies validate that Re4 helps to learn learning distinct and effective multi-interest representations.
\end{abstract}




\maketitle


\section{Introduction}

A proliferation of the Internet has resulted in an increase in information overload in people's daily lives. Recommender systems help users seek desired information and explore what they are potentially interested in, thus alleviating the information overload. It has become one of the most widely used information systems with applications such as E-commerce, news portals, and micro-video platforms. A successful recommendation framework depends on accurately describing and representing users' interests. The recent advances in neural recommender systems have convincingly demonstrated high capability in learning dense user representation for matching. Matching (also known as deep candidate generation) methods typically represent users and items with dense vectors, and leverage simple similarity functions (\eg, dot product, and cosine similarity) to model user-item interactions. Typically, YoutubeDNN \cite{Covington_Adams_Sargin_2016} takes user behavior sequence as input, perform mean-pooling on item embeddings in the sequence to obtain user embedding\footnotetext[1]{These authors contributed equally to this work.}\footnotetext[2]{Corresponding Authors.}. 

Despite their effectiveness, most existing works typically represent each user using an overall embedding. However, in real-world applications, users' interests exhibit multiple aspects. For example, in E-commerce platforms, a user might be simultaneously in favor of sports equipment (\eg, basketball) and electronic products (\eg, desktop). In the embedding hypersphere, an overall user embedding might be less effective in capturing multiple item clusters. As such, devising multi-interest representation frameworks is a promising research direction for capturing users' diverse interests. Multi-interest recommendation is still a nascent research area. Recently, MIND \cite{Li_Liu_Wu_Xu_Zhao_Huang_Kang_Chen_Li_Lee_2019} groups users’ historical behaviors into multiple clusters based on dynamic capsule routing while each interest capsule reflects a particular aspect. ComiRec \cite{Cen_Zhang_Zou_Zhou_Yang_Tang_2020} introduces self-attention mechanisms to extract multiple interest embeddings and a controllable factor to realize the diversity-performance tradeoff.

However, existing methods only rely on the \textit{forward flow} (items to multi-interest), but do not consider the information from multi-interest to items, named as backward flow. Ignoring the backward flow might raise some limitations on the learned embeddings. For example, a representative attention-based multi-interest framework ComiRec-SA extracts multiple interests using different attention heads. The different heads of attention mechanism mainly introduce randomness into the modeling process but do not necessarily guarantee the outputs of different heads to be distinct from each other. As a result, there is no guarantee on whether the learned multiple embeddings represent multiple aspects of interest.
Meanwhile, the attention weights are interpreted as the correlation between historical items and different interests. However, there is no guarantee on whether the correlation is consistent with the recommendation criterion on user-item matching. Note that we use the latter criterion for the final recommendation serving.
Therefore, an effective and robust multi-interest recommendation framework requires backward flows (interests-to-items) to re-examine the relationships of learned multi-interest embeddings and historical items. 

To address the aforementioned challenges, we propose to leverage the \textit{backward flow} (multi-interests to items) and construct the \textbf{Re4} framework for multi-interest recommendation. Re4 consists of three essential components, \ie, \textit{Re-contrast}, \textit{Re-attend}, and \textit{Re-construct}, for effective multi-interest \textit{Re}commendation. 1) \textbf{Re-contrast} is devised to learn multiple distinct interests that should capture different aspects of users' interests. Technically, for each interest, we firstly identify the representative items corresponding to the interest based on the similarity function used for deep candidate generation, such as dot product or cosine similarity. Then, we view the representative items as positive items for the interest. There are three kinds of negatives, \ie, items in the user sequence except for the positives, randomly sampled items beyond the sequence, and other interests.
Finally, we conduct contrastive learning by pulling the interest closer to the positives and pushing the interest away from the negatives. Obviously, this modeling drives each interest to be distinct from the others. 2) \textbf{Re-attend} addresses the concern on the consistency between the attention weights in the forward flow and the recommendation correlation between interests and users. The attention weights are used as the basis for item clustering and interest learning. The correlation is used for final recommendation. Therefore, ensuring consistency between them is essential. Technically, Re-attend explicitly minimizes the distance between the forward attention weights and the interest-item correlation. 3) the above two strategies focus on correlation, \ie, to what extent each interest and each historical item is correlated. As a counterpart, we leverage \textbf{Re-attend} to ensure each interest representation can semantically reflect the content of representative items.

We conduct extensive experiments on three public benchmarks and demonstrate the effectiveness of Re4 against state-of-the-art multi-interest frameworks on the matching phase of recommendation. Various analysis including ablation study, hyper-parameter analysis, and case study validate the practical merits of Re4 on learning effective multi-interest representations. In summary, the main contributions of this work are threefold:
\begin{itemize}[leftmargin=*]
	\item We make an early attempt to incorporate backward flow (interests-to-items) modeling for multi-interest representation learning. 
	\item We propose the Re4 framework and devise three backward flows, \ie, Re-contrast, Re-attend, and Re-construct, to learn distinct multi-interests that can semantically reflect representative items.
	\item We conduct extensive experiments on three real-world datasets, validating the effectiveness of Re4 on multi-interest recommendation. 
\end{itemize}

\begin{figure}[!t] \begin{center}
    \includegraphics[width=\columnwidth]{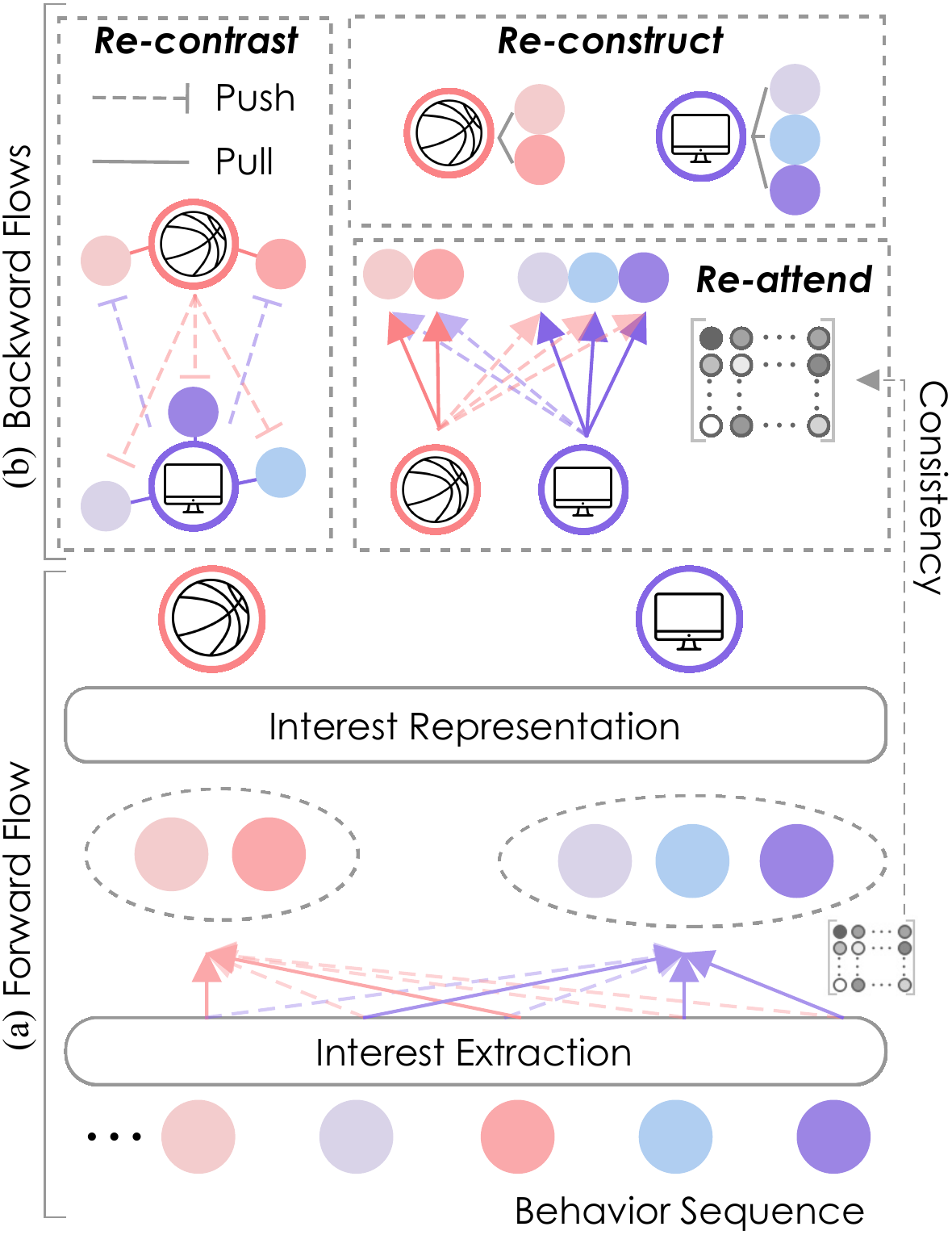}
    \caption{
    	An illustration of leveraging backward flows for multi-interest representation learning. (a) The traditional forward flow that clusters items and extracts multiple interests. (b) The proposed backward flows, \ie, \textit{Re-contrast} which learns distinct multi-interests; \textit{Re-construct} which permits interests' semantic reflection on representative items; and \textit{Re-attend} which ensures the consistency between attention weights in the forward flow and recommendation correlation.
	}
\label{fig:longtail_base}
\end{center} \end{figure}

\section{Methods}

In this section, we will elaborate on the building blocks of Re4, \ie, the multi-interest extraction module, which is the forward flow, and three backward flow strategies. We use bold letters ($\eg, \mathbf{u}$) to denote vectors, bold upper-case letters ($\eg, \mathbf{W}$) to denote matrices, and letters in calligraphy font ($\eg, \mathbf{P}$) to denote sets.

\subsection{Multi-interest Extraction} \label{sec:mie}

Given a behavior sequence $X = \{ x^u_i \}_{i = 1,\dots,N_x}$ of user $u$, the multi-interest extraction module aims to extract multiple dense vectors $\mathbf{Z}^u \in \mathbb{R}^{d \times N_z} = \{ \mathbf{z}_k^u \}_{k=1,\dots,N_z}$. $x^u_i$ denotes the $i$th behavior of user $u$, and $N_x$ is the length of the behavior sequence. $\mathbf{z}_k^u$ represents the $k$th interest embedding of user $u$, and $N_z$ is a hyper-parameter indicating the number of interests. For simplicity, we will drop the superscripts occasionally and use $x_i$ and $\mathbf{z}_k$ in place of $x_i^u$ and $\mathbf{z}_k^u$.

Currently, there are two widely used approaches for this aim, \ie, dynamic capsule routing \cite{Li_Liu_Wu_Xu_Zhao_Huang_Kang_Chen_Li_Lee_2019} and attention mechanisms \cite{Cen_Zhang_Zou_Zhou_Yang_Tang_2020}.
We leverage attention mechanisms due to their effectiveness in a broad range of deep learning applications.
Specifically, we first transform the behavior sequence into the dense representation using trainable item embedding table, \ie, $\mathbf{X} \in \mathbb{R}^{N_x \times d} = \{ \mathbf{x}_i \}_{i = 1,\dots,N_x}$. Then, we employ the additive attention technique to obtain attention weight of the $k$th interest on the $i$th item:
\begin{align}
	a_{k,i} = \frac{\exp \left( \mathbf{w}_k^T \tanh \left( \mathbf{W}_1 \mathbf{x}_i \right) \right)}{ \sum_j \exp \left( \mathbf{w}_k^T \tanh \left( \mathbf{W}_1 \mathbf{x}_j \right) \right) }, \label{eq:mi1}
\end{align}
where $\mathbf{W}_1 \in \mathbb{R}^{d_h \times d}$ is a transformation matrix shared by all interests, and $\mathbf{w}_k \in \mathbb{R}^{d_h}$ is a interest-specific transformation vector to compute interest-item correlation. $a_{k,i \in \mathbb{R}}$ indicates to what extent item $x_i$ belongs to the interest $z_k$. The $k$th interest representation is obtained by:
\begin{align}
	\mathbf{z}_k = \sum_j a_{k,j} \mathbf{W}_2 \mathbf{x}_j. \label{eq:mi2}
\end{align}

\subsection{Backward Flow} 

The multi-interest extraction module solely models the item-to-interest forward flow. We argue that interest-to-item backward flow can further enhance multi-interest representation learning. Specifically, we devise three backward flows and elaborate their details as the following.

\subsubsection{Re-contrast.} As shown in Equation \ref{eq:mi1}-\ref{eq:mi2}, the attention mechanism extracts multiple interests with interest-specific transformation parameters $\mathbf{w}_k$. However, there is no guarantee that neither $\mathbf{w}_k$ nor the attention weights to be different for different interests. Each $\mathbf{w}_k$ can be interpreted as an attention head, and different attention heads are known to introduce randomness rather than diversity. Therefore, there are chances that the model learns a trivial solution where all interests are close in the embedding space \wrt items in the behavior sequence. Towards this end, we construct the Re-contrast backward flow, which leverages contrastive learning \cite{Oord_Li_Vinyals_2018,Wu_Xiong_Yu_Lin_2018} to learn distinct interest representations. Basically, contrastive learning is performed by pushing the anchor away from the negatives and pulling the anchor closer to the positives. Obviously, the essence of contrastive learning lies in the construction of effective positives and negatives.

\vpara{Positives.} Undoubtedly, the representative items corresponding to the anchor interest can be viewed as positives. In our multi-interest framework, the items in the behavior sequence with high attention weights can be interpreted as representative items. As such, we perform hard selection as the following:
\begin{align}
	\mathcal{P}_k = \{ \mathbf{x}_j \mid a_{k, j} > \gamma_c  \}, \label{eq:pos}
\end{align}
where $\mathcal{P}_k$ denotes the set of positives which includes items with attention weight  $a_{k, j}$ higher than a certain threshold $\gamma_c$. We empirically set the threshold to the uniform probability $1 / N_x$.

\vpara{Negatives.} As a counterpart of the above positives, a straightforward solution is to view other items in the behavior sequence as negatives:
\begin{align}
	\mathcal{\bar N}_k = \{ \mathbf{x}_j \mid a_{k, j} \leq \gamma_c  \},
\end{align}
However, note that our primary goal is to diversify learned interests. The above learning can make each interest representation more discriminative \wrt items that are not representative, but not necessarily make each interest different from each other. Therefore, we propose to add other interests as negatives. Moreover, we add items that do not appear in the behavior sequence to stabilize contrastive learning. The final negative set can be obtained by:
 \begin{align}
	\mathcal{N}_k = \mathcal{\bar N}_k \cup \left(\mathcal{Z} \setminus \{ \mathbf{z}_k \} \right) \cup \mathcal{\tilde N}_k,
\end{align}
where $\mathcal{Z}$ denotes the set of interest representations, and $\mathcal{\tilde N}_k$ is the set of items representations that are sampled beyond the behavior sequence. With the positives $\mathbf{z}_{k,i}^+ \in \mathcal{P}_k$ and negatives $\mathbf{z}_{k,j}^- \in \mathcal{N}_k$, we employ InfoNCE \cite{Oord_Li_Vinyals_2018} as the objective:
\begin{align}
	\mathcal{L}_{CL} = - \sum_i log \frac{\exp \left( \mathbf{\bar z}_k \cdot \mathbf{\bar z}^+_{k,i} / \tau \right)}{ \exp \left( \mathbf{\bar z}_k \cdot \mathbf{\bar z}^+_{k,i} / \tau \right) + \sum_j \exp \left( \mathbf{\bar z}_k \cdot \mathbf{\bar z}^-_{k,j} / \tau \right) }.
\end{align} 
where $(\mathbf{\bar z}_k, \mathbf{\bar z}^+_{k,i}, \mathbf{\bar z}^-_{k,j})$ are L2 normalized vectors of $(\mathbf{z}_k, \mathbf{z}^+_{k,i}, \mathbf{z}^-_{k,j})$, and $\tau$ is temperature hyper-parameter \cite{Wu_Xiong_Yu_Lin_2018}.

\begin{figure}[!t] \begin{center}
    \includegraphics[width=\columnwidth]{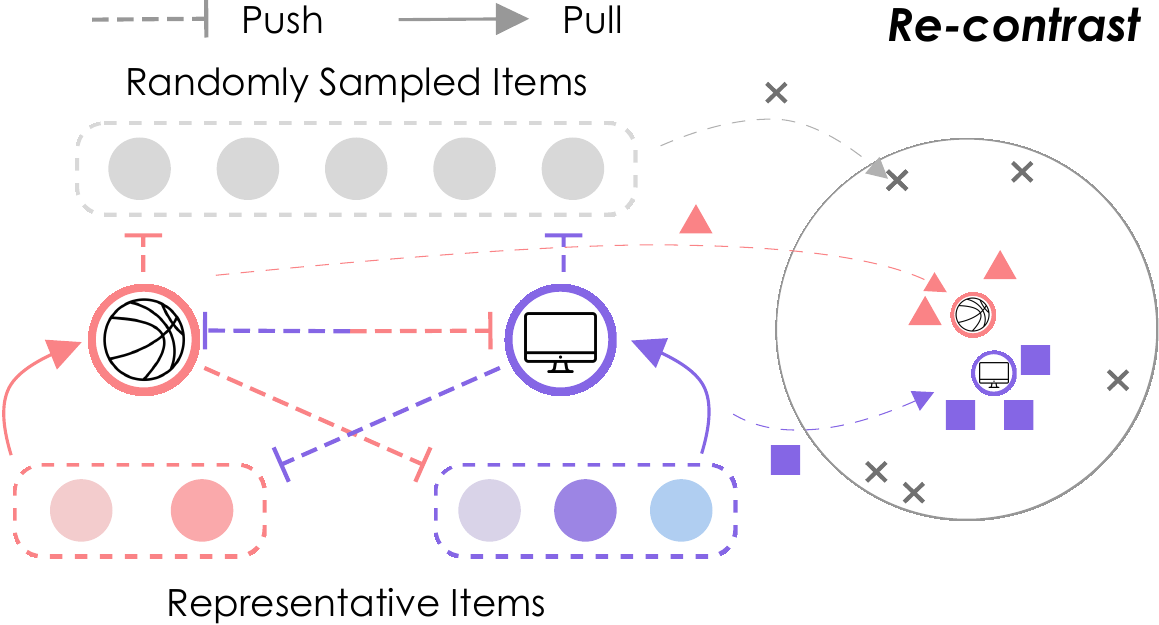}
    \caption{
    	Schema of the Re-contrast backward flow, which aims to learn distinct multi-interest representations.
	}
\label{fig:longtail_base}
\end{center} \end{figure}

\subsubsection{Re-attend} In the multi-interest extraction module, the attention weight $a_{k,i}$ is interpreted as the correlation between the $k$th interest and the $i$th item in the behavior sequence. However, such a correlation is not consistent with how matching models make recommendations. Typically, a matching model leverages dot product or cosine similarity to estimate the probability of users clicking on items. As such, to make the correlation computation in the forward flow consistent with the correlation measurement in the final recommendation, we construct the Re-attend backward flow. We first compute the correlation between interests and historical items using the recommendation measurement $\phi$:
\begin{align}
	{\tilde a}_{k,i} = \phi(\mathbf{z}_k, \mathbf{x}_i),
\end{align}
where $\phi$ is determined according to the recommender system and we use dot product in experiments.
The Re-attend loss function can be written as:
\begin{align}
	\mathcal{L}_{Att} = \sum_k \sum_i L_{CE} ({a}_{k,i}, {\tilde a}_{k,i}).
\end{align}
where $L_{CE}$ denotes the cross-entropy loss function.

\subsubsection{Re-construct} The above two backward flows are concerned with correlations, \ie, to what extent interest-interest and interest-item are correlated. However, they neglect whether interest representations can reflect the content of representative items. To permit such a semantic reflection, we construct the Re-construct backward flow. We leverage self-attention mechanism for reconstruction, which is formulated as:
\begin{align}
	\mathbf{C}_k &= \operatorname{Upsample} \left( \mathbf{z}_k \right), \\
	\beta_{k,i,j} &= \frac{\exp \left( \mathbf{\bar w}_j^T \tanh \left( \mathbf{\bar W}_3 \mathbf{c}_{k,i} \right) \right)}{ \sum_m \exp \left( \mathbf{\bar w}_j^T \tanh \left( \mathbf{\bar W}_3 \mathbf{c}_{k,m} \right) \right) }, \\
	\mathbf{\hat x}_{k,j} &= \sum_i \beta_{k,i,j} \mathbf{\bar W}_5 \mathbf{c}_{k,i},
\end{align}
where the $\operatorname{Upsample}$ function is a linear projection $\mathbf{\bar W}_4 \in \mathbb{R}^{N_xd_b \times d}$ followed by a reshape operation to transform the linearly projected vector to a matrix $\mathbf{C}_k \in \mathbb{R}^{N_x \times d_b}$. $d_b$ is the hidden size in Re-construct backward flow. $\mathbf{c}_{k,i}$ is the $i$th unit in $\mathbf{C}_k$. $\mathbf{\bar W}_{3} \in \mathbb{R}^{d_b \times d_b}$, $\mathbf{\bar W}_{5} \in \mathbb{R}^{d \times d_b}$, and $\mathbf{\bar w}_{j} \in \mathbb{R}^{d_b}$ are learnable transformations, and each input item $x_j$ has a corresponding $\mathbf{\bar w}_{j}$.
Different from auto-encoders \cite{Sedhain_Menon_Sanner_Xie_2015,Liang_Krishnan_Hoffman_Jebara_2018,Khawar_Poon_Zhang_2020} which reconstruct all inputs, we propose to reconstruct representative items corresponding to the interest. Specifically, we take the positive set $\mathcal{P}_k$ constructed by Equation \ref{eq:pos} as the representative items for the $k$th interest. Therefore, the loss function of the Re-construct backward flow can be written as:
\begin{align}
	\mathcal{L}_{CT} = \sum_k \sum_j \mathbbm{1}\left( \mathbf{x}_j \in \mathbb{P}_k \right) \norm{ \mathbf{\hat x}_{k,j} - \mathbf{x}_{j} }^2_F.
\end{align}
where $\mathbbm{1}$ is an indicator function which returns 1 when the condition is true and returns 0 otherwise.
We empirically find that semantic reflection can help the interest presentation to have a fine-grained understanding of items, and thus leads to boost recommendation metrics that consider rank positions.

\begin{table}[!t]
\caption{Dataset Statistics.}
\centering
\begin{tabular}{l cc cc}
\toprule
Dataset & \#Users & \#Items & \#Interactions & \#Density  \\
    \midrule
    Book          & 603, 668 & 367, 982 & 8, 898, 041 & 0.00004 \\ 
    MovieLens   & 6, 040 & 3, 707 & 1, 000, 209 & 0.04467 \\
    Gowalla   & 29, 858 & 40, 981 & 1, 027, 370 & 0.00084 \\
    \bottomrule
\end{tabular}%
    \label{tab:staData}
\end{table}
%

\subsection{Training and Inference}

We follow the training and inference paradigm of ComiRec \cite{Cen_Zhang_Zou_Zhou_Yang_Tang_2020}. At the training stage, given multiple interest representations $\mathbf{Z} = \{ \mathbf{z}_k \}_{k=1,\dots,N_z}$ and the target item embedding $\mathbf{y}$, we obtain the interest embedding that is the most related to the target item through:
\begin{align}
	\mathbf{\hat z} = \mathbf{Z} \left[:, \operatorname{argmax} \left( \mathbf{Z}^T \mathbf{y} \right) \right],
\end{align}
Then, we adopt the negative log-likelihood objective:
\begin{align}
	\mathcal{L}_{Rec} &= -\log p_\theta(y \mid X), \label{eq:sim}
    \quad \\
    \mathrm{where}\; p_\theta(y \mid X) &= \frac{\exp \left(  \mathbf{\hat z}^T\mathbf{y}  \right) }{\sum_{y^\prime \in \mathcal{Y}}{\exp \left( \mathbf{\hat z}^T\mathbf{y}^\prime \right) }},
    \label{eq:mle}
\end{align}
where $y^\prime$ denotes a randomly sampled item. During the matching phase of recommendation, it can be impractical to sum over the entire item gallery $\mathcal{Y}$ as in the denominator. We adopt the sample softmax objective \cite{Bengio_Senecal_2008,Jean_Cho_Memisevic_Bengio_2015}. Therefore, the final training loss function is:
\begin{align}
	\mathcal{L}_{Re4} = \mathcal{L}_{Rec} + \lambda_{CL} \mathcal{L}_{CL} + \lambda_{Att} \mathcal{L}_{Att} + \lambda_{CT} \mathcal{L}_{CT}.
\end{align}

At inference, the recommendation probability of item $y$ for user $u$ with multiple interests $\mathbf{Z}^u = \{ \mathbf{z}_k^u \}_{k=1,\dots,N_z}$ is:
\begin{align}
	p_{u,y} = \max \left\{ \mathbf{y}^T \mathbf{z}_{k}^u  \right\}_{k=1,\dots,N_z}
\end{align}
Finally, the top-N items are obtained with $p_{u,y}$ as the basis.

\begin{table*}[h]
\centering
    \caption{ A comparison between the proposed Re4 framework and state-of-the-art matching baselines on three public benchmarks. Re4 mostly achieves performance gains over baselines, and the performance improvement is more substantial on the larger-scale dataset with a large item gallery (Amazon) where users' interests are more likely to be diverse.}
{\setlength{\tabcolsep}{0.8em}\renewcommand{\arraystretch}{1.0}\begin{tabular}{ll cccccccc cc}
\toprule

  Datasets  & Metric & POP & Y-DNN  & GRU4Rec &  MIND & ComiRec-SA & ComiRec-DR  & Re4  & Improv.   \\
    \midrule
   
  \multirow{6}{*}{Amazon} 
   &  R@20  &  0.0137  &  0.0457   &  0.0406    & 0.0486  &  \underline{0.0549}  &  0.0531  &  \textbf{0.0771}  & 40.44\%  \\
   &  R@50  &  0.0240  &  0.0731   &  0.0650    & 0.0764  &  \underline{0.0847}  &   0.0811   &  \textbf{0.1155}   & 36.36\%  \\
   &  NDCG@20  &  0.0226  &  0.0767   &  0.0680   & 0.0793   &  0.0899 & \underline{0.0918}    &  \textbf{0.1304}  &  42.05\% \\
   &  NDCG@50  &  0.0394  &  0.1208   &   0.1037   & 0.1223  &  \underline{0.1356}  & 0.1352   &  \textbf{0.1883}  & 38.86\%  \\
   &  HR@20  & 0.0302   &  0.1029   &   0.0894   &  0.1062 &  0.1140  &\underline{0.1201}    & \textbf{0.1627}   &  35.47\%  \\
   &  HR@50  & 0.0523   &  0.1589   &   0.1370   &  0.1610 &  0.1720  & \underline{0.1758}   &  \textbf{0.2326}  & 32.31\%  \\
    \midrule

    \multirow{6}{*}{MovieLens} 
   &  R@20  	&  	0.0006  & 	0.1115   &  \textbf{0.1286}	 &    0.1033	   &   0.1189  &   \underline{0.1223}     & 0.1117   &  -13.14\% \\
   &  R@50  	&  	0.0016  & 	0.2191   &  \textbf{0.2428}	 &    0.1994	   &   0.1949  &   \underline{0.2263}     & 0.2048   &  -15.65\% \\
   &  NDCG@20  	&  	0.0057  & 	0.3671   &  \underline{0.3971}	 &    0.3325	   &   0.3131  &   0.3913     & \textbf{0.4581}   &  15.36\% \\
   &  NDCG@50  	&  	0.0135  & 	0.4035   &  \underline{0.4157}	 &    0.3683	   &   0.3396  &   0.4039     & \textbf{0.6067}   &  45.95\% \\
   &  HR@20  	&  	0.0186  & 	0.7318   &  \underline{0.7831}	 &    0.7020	   &   0.7550  &   0.7714     & \textbf{0.8048}   &  2.77\% \\
   &  HR@50  	&  	0.0452  & 	0.8858   &  \underline{0.8990}	 &    0.8593	   &   0.8874  &   0.8801     & \textbf{0.9288}   &  3.31\% \\
    \midrule
    \multirow{6}{*}{Gowalla} 

   &  R@20  	&  	0.0028  & 	0.1127   &  0.1273	 &    0.1218	   &   \underline{0.1277}  &   0.1153     & \textbf{0.1386}   &  8.54\% \\
   &  R@50  	&  	0.0054  &  	0.1926  & 0.2043 	 &  	 0.2049   & \underline{0.2072}    &  0.1831      & \textbf{0.2203}    & 6.32\%  \\
   &  NDCG@20 	 &  0.0073	  & 	0.2378   &  \underline{0.2803}	&    0.2565  	     & 0.2736    &   0.2534    &  \textbf{0.3141}  & 12.06\%  \\
   &  NDCG@50  	&   0.0135	 &  	0.3638       & 0.4002     & 	      0.3888    &  \underline{0.4019}   &   0.3621   &  \textbf{0.4412}   & 9.78\%  \\
   &  HR@20 	 &  	0.0104  &  	0.3443  & 0.3814 	&     0.3627   	    &  \underline{0.3838}   &   0.3429     &  \textbf{0.4206}  &  9.59\% \\
   &  HR@50 	 &  	0.0224  & 	0.5010   &	 0.5251 &   0.5301       &  0.5288   &   \underline{0.5355}    &  \textbf{0.5697}  & 6.39\%  \\

    \bottomrule
\end{tabular}}
    \label{tab:comparison}
\end{table*}
%

\section{Related Works}

\vpara{Neural Recommender Systems.} Neural recommendation models incorporate neural networks for user-item interaction modeling or user/item representation learning. Neural networks are graceful to capture the non-linear feature interactions between users and items~\cite{hsieh2017collaborative,he2017neural,wu2016collaborative,guo2017deepfm,Zhao_Lu_Cai_He_Zhuang_2016,wang2021clicks,wang2021deconfounded,Yao_Wang_Jia_Han_Zhou_Yang_2021,Yao_Zhang_Yao_Wang_Ma_Zhang_Chu_Ji_Jia_Shen_2021}, and can hereby boost traditional collaborative filtering methods. Typically, neural collaborative filtering (NCF) \cite{He_Liao_Zhang_Nie_Hu_Chua_2017} leverages both a generic matrix factorization component and a non-linear MLP for interaction modeling to jointly enhance recommendation. As for user/item representation learning, there are roughly two lines of works, \ie, graph-based modeling \cite{He_Deng_Wang_Li_Zhang_Wang_2020,Berg_Kipf_Welling_2017,Wang_He_Wang_Feng_Chua_2019,Zheng_Lu_Jiang_Zhang_Yu_2018,Sun_Guo_Zhang_Zhang_Regol_Hu_Guo_Tang_Yuan_He_et,Wang_Jin_Zhang_He_Xu_Chua_2020}, sequence-based modeling \cite{Manotumruksa_Yilmaz_2020,Ren_Liu_Li_Zhao_Wang_Ding_Wen_2020,Wang_Zhang_Ma_Liu_Ma_2020,Yuan_He_Karatzoglou_Zhang_2020,Ma_Ren_Lin_Chen_Ma_Rijke_2019,Wang_Guo_Lan_Xu_Wan_Cheng_2015,Zhang_Yao_Zhao_Chua_Wu_2021,Lu_Zhang_Huang_Wang_Yu_Zhao_Wu_2021,Xun_Zhang_Zhao_Zhu_Zhang_Li_He_He_Chua_Wu_2021}. Sequence-based recommendation models have the advantage of modeling dynamic user interest by extracting user representation from the newest behavior sequence. Youtube-DNN \cite{Covington_Adams_Sargin_2016} is one of the earliest works on sequential recommendation which leverages mean-pooling to obtain users' representation. Inspired by sequence modeling in the generic domain (\eg, natural language processing and video processing~\cite{ye2022,DBLP:journals/corr/abs-2104-07650,Zhao_Xiao_Song_Lu_Xiao_Zhuang_2020,Zhao_Zhang_Xiao_Xiao_Yan_Yu_Cai_Wu_2019,Zhao_Zhang_Jiang_Cai_2019,Zhang_Jiang_Wang_Kuang_Zhao_Zhu_Yu_Yang_Wu_2020,Zhang_Tan_Yu_Zhao_Kuang_Liu_Zhou_Yang_Wu_2020,Zhang_Tan_Zhao_Yu_Kuang_Jiang_Zhou_Yang_Wu_2020,zhang2021consensus,zhang2020relational,zhang2021magic}), this work is followed by a lot of advanced techniques such as Recurrent Neural Networks \cite{Hidasi_Karatzoglou_Baltrunas_Tikk_2016,Donkers_Loepp_Ziegler_2017,Hidasi_Karatzoglou_2018,Li_Ren_Chen_Ren_Lian_Ma_2017,Quadrana_Karatzoglou_Hidasi_Cremonesi_2017,Yu_Liu_Wu_Wang_Tan_2016}, attention mechanisms \cite{Wang_Hu_Cao_Huang_Lian_Liu_2018,Ying_Zhuang_Zhang_Liu_Xu_Xie_Xiong_Wu_2018,Sun_Liu_Wu_Pei_Lin_Ou_Jiang_2019,Cen_Zhang_Zou_Zhou_Yang_Tang_2020}, dynamic capsule routing~\cite{Cen_Zhang_Zou_Zhou_Yang_Tang_2020,Li_Liu_Wu_Xu_Zhao_Huang_Kang_Chen_Li_Lee_2019}, and memory networks \cite{Chen_Xu_Zhang_Tang_Cao_Qin_Zha_2018,Huang_Zhao_Dou_Wen_Chang_2018}.

\vpara{Multi-interest Recommendation.} To capture diverse interests of users, there is increasing attention on multi-interest representation learning \cite{Chen_Liu_Xiong_Zha_2021,Chen_Zhang_Zhao_Xue_Xiang_2021,Wu_Yin_Lian_Yin_Gong_Zhou_Yang_2021,Cen_Zhang_Zou_Zhou_Yang_Tang_2020,Li_Liu_Wu_Xu_Zhao_Huang_Kang_Chen_Li_Lee_2019,Xiao_Yang_Jiang_Wei_Hu_Wang_2020}, related to disentangled representation in the generic domain~\cite{wu2020learning,li2020ib}. MIND \cite{Li_Liu_Wu_Xu_Zhao_Huang_Kang_Chen_Li_Lee_2019} takes the initiative to represent users with multiple interests. They devise the multi-interest extractor based on dynamic capsule routing \cite{Sabour_Frosst_Hinton_2017}. ComiRec \cite{Cen_Zhang_Zou_Zhou_Yang_Tang_2020} is a state-of-the-art that leverages self-attention for multi-interest modeling. They also introduce a controllable factor for recommendation diversity-accuracy tradeoffs. We follow these works to devise generic multi-interest modeling for the matching phase, and propose to model the backward flow (interests-to-items). \cite{Chen_Zhang_Zhao_Xue_Xiang_2021} utilizes auxiliary time information to better extract multiple interests. \cite{Chen_Liu_Xiong_Zha_2021} construct pre-defined four kinds of representation (user-level, item-level, neighbor-assisted, and category-level) for video recommendation. \cite{Xiao_Yang_Jiang_Wei_Hu_Wang_2020} focuses on the ranking phase of recommendation, and devises a target-item-aware multi-interest extraction layer.

\vpara{Backward Flow in Recommendation.} Generally, leveraging backward flow (output-to-input) is not a new paradigm for deep learning. Auto-encoders \cite{Sedhain_Menon_Sanner_Xie_2015,Wu_DuBois_Zheng_Ester_2016,Liang_Krishnan_Hoffman_Jebara_2018,Sachdeva_Manco_Ritacco_Pudi_2019,Khawar_Poon_Zhang_2020} and dual learning \cite{He_Xia_Qin_Wang_Yu_Liu_Ma_2016,Lee_Park_Lee_2021,Zhang_Cheng_Yao_Yi_Hong_Chi_2021,Sun_Wu_Zhang_Fu_Hong_Wang_2020,Zhuang_Zhang_Qian_Shi_Xie_He_2017} share similar ideas. \cite{Khawar_Poon_Zhang_2020} removes unnecessary connections between neurons in existing fully-connected auto-encoders and enhances recommendation based on the optimized encoder structures. \cite{Sun_Wu_Zhang_Fu_Hong_Wang_2020} views preference prediction and review generation as two dual tasks, and interchangeably predicts each one using the other along with user and item id information. To the best of our knowledge,  there is no prior attempt to explore the backward flow in multi-interest recommendation. Moreover, we propose several practical strategies that boost multi-interest representation learning in semantics and in correlation with corresponding representative items.

\section{Experiments}

\subsection{Experimental Setup}

We conduct experiments on real-world datasets to answer three main research questions:

\begin{itemize}[leftmargin=*,labelsep=0.8mm]
\renewcommand\labelitemi{{\boldmath$\cdot$}}
	\item \textbf{RQ1}: How does Re4 perform compared to state-of-the-art models?
	\item \textbf{RQ2}: How do the different re-examination strategies and the number of interests affect Re4?
	\item \textbf{RQ3}: How do the learned multi-interest representations benefit from the backward flow?
\end{itemize}

\vpara{Datasets} We consider three real-world recommendation benchmarks, of which the statistics are shown in Table \ref{tab:staData}.
\begin{itemize}[leftmargin=*,labelsep=0.8mm]
\renewcommand\labelitemi{{\boldmath$\cdot$}}
\item \textbf{Amazon\footnote{http://jmcauley.ucsd.edu/data/amazon/}.} Amazon Review dataset \cite{McAuley_Targett_Shi_Hengel_2015} is one of the most widely used recommendation benchmarks. We choose the largest category Amazon Book for evaluation. We keep the last 20 behaviors to construct the behavior sequence for each user.
\item \textbf{MovieLens\footnote{https://grouplens.org/datasets/movielens/1m/}.} MovieLens-1M is a widely used public benchmark on movie ratings.
\item \textbf{Gowalla.} We use the 10-core setting~\cite{He_McAuley_2016} of the check-in dataset~\cite{Liang_Charlin_McInerney_Blei_2016} released by the Gowalla platform.
\end{itemize}

\vpara{Evaluation Protocal \& Metrics.} For a fair comparison, we employ the evaluation framework of ComiRec \cite{Cen_Zhang_Zou_Zhou_Yang_Tang_2020}, which can demonstrate the generalization capability of all models by assessing them according to unseen user behavior sequences. Details can be found in \cite{Cen_Zhang_Zou_Zhou_Yang_Tang_2020}.
The \textbf{\textit{matching}} phase of recommendation is our primary focus, and we select the datasets, comparison methods, and evaluation metrics accordingly.
For the matching phase, \textit{Recall}, \textit{Normalized Discounted Cumulative Gain} (NDCG)\footnote{For a fair comparison with the state-of-the-art baseline ComiRec, we strictly follow their metric implementation. The normalization term IDCG used in their paper is calculated \wrt recalled positive items rather than all positive items.}, and \textit{Hit Rate} are three broadly used metrics. 
Metrics are computed based on the top 20/50 recommended candidates (\eg, Recall@20). Higher scores demonstrate better recommendation performance for all metrics.

\vpara{Baselines} We follow ComiRec \cite{Cen_Zhang_Zou_Zhou_Yang_Tang_2020} to construct matching baselines and also take their method for comparison. Since the evaluation protocol requires modeling unseen users and unseen behavior sequences, we do not consider factorization-based and graph-based methods.
\begin{itemize}[leftmargin=*,labelsep=0.8mm]
	\renewcommand\labelitemi{{\boldmath$\cdot$}}
	\item \textbf{POP.} POP always recommends the most popular items to users.
	\item \textbf{YouTube DNN\cite{Covington_Adams_Sargin_2016}.} Y-DNN is a successful industrial recommender that takes the behavior sequence as input, and pools the embedded items to have user representation.
	\item \textbf{GRU4Rec \cite{Hidasi_Karatzoglou_Baltrunas_Tikk_2016}.} GRU4Rec is a representative model that uses the Gated Recurrent Unit \cite{Cho_Merrienboer_Glehre_Bahdanau_Bougares_Schwenk_Bengio_2014} to model the sequential dependencies between items.
	\item \textbf{MIND \cite{Li_Liu_Wu_Xu_Zhao_Huang_Kang_Chen_Li_Lee_2019}.} MIND is one of the earliest multi-interest frameworks, and employs dynamic capsule routing to extract multiple interest embeddings.
	\item \textbf{ComiRec-DR \cite{Cen_Zhang_Zou_Zhou_Yang_Tang_2020}.}  ComiRec-DR is the state-of-the-art multi-interest framework that also uses dynamic routing and introduces a controllable factor for recommendation diversity-accuracy tradeoffs.
	\item \textbf{ComiRec-SA \cite{Cen_Zhang_Zou_Zhou_Yang_Tang_2020}.} ComiRec-SA uses self-attention mechanisms for multi-interest modeling, which is the base model of Re4.
\end{itemize}

\vpara{Implementation Details} 
We use Adam \cite{Kingma_Ba_2015} for optimization with learning rate of 0.003/0.005 for Books/Yelp and Gowalla, $\beta_1= 0.9$, $\beta_2 = 0.99$, $\epsilon=1 \times 10^{-8}$, weight decay of $1 \times 1e-5$. All models are with embedding size $d=64$. The hidden size in the forward flow is set to $d_h=256$, and the hidden size in the backward flow is set to $d_b = 32$. The temperature in Re-contrast is set to $\tau=0.02$. The default interest number in experiments is set to $N_z=8$. We search for loss coefficients $\lambda_{Att}$, $\lambda_{Att}$, and $\lambda_{Att}$ in the range $\{ 0.01, 0.1, 1, 10 \}$. Re4 is implemented with Pytorch \cite{NEURIPS2019_9015} 1.6.0 and Python 3.8.5.

\subsection{Overall Performance (RQ1)} \label{sec:overall}

Table \ref{tab:comparison} shows the recommendation results on three datasets. We have the following observations:

\begin{itemize}[leftmargin=*,labelsep=0.8mm]
\renewcommand\labelitemi{{\boldmath$\cdot$}}
	\item Always recommending popular items (POP) without considering users' interests is less effective on three datasets. These results show that neural models cannot easily achieve better results by fitting trivial findings (\eg, recommending items that have more interactions).
	\item Traditional sequential recommenders (Y-DNN and GRU4Rec) substantially outperform POP. Sequential recommenders have the advantage of taking users' latest behavior sequence into consideration, and can hereby make dynamic recommendations for users with interactions that are unseen during training. GRU4Rec consistently outperforms Y-DNN, demonstrating the superiority of modeling sequential dependencies between items. Note that Y-DNN solely mean-pools historical items. However, both Y-DNN and GRU4Rec represent users with an overall user embedding, which might easily lead to suboptimal results when items are various, and users' interests are diverse.
	\item Multi-interest frameworks (\eg, MIND, ComiRec-SA) generally achieve better performance than baselines that uses single embedding to represent users on large-scale datasets (\ie, Amazon, and Gowalla). Such improvement basically indicates that when items are diversified and users' interests are hereby more likely to be diverse, multi-interest framework is a more effective way to represent users. Not surprisingly, on MovieLens-1M with limited items (3, 900), multi-interest baselines cannot beat traditional recommenders. 
		Among multi-interest baselines, ComiRec-SA with self-attention mechanisms achieves the best performance results on three datasets. It makes sense that attention mechanisms have been demonstrated as effective in numerous deep learning tasks \cite{vaswani2017attention,Cheng_Dong_Lapata_2016,Lu_Batra_Parikh_Lee_2019}. The different attention heads in ComiRec-SA introduce randomness and are interpreted as interest encoders for different interest aspects. The attention weights of each interest on historical items are hereby interpreted as the correlation of items and interests.
	\item Re4 consistently yields the best performance on three datasets in most cases. Remarkably, Re4 improves the best performing baselines by 42.05\%, 15.36\%, and 12.06\% in terms of NDCG@20 on Amazon Book, MovieLens, and Gowalla, respectively. By leveraging the backward flows, the learned multi-interests become more distinct from each other, and better correlated with the corresponding representative items. Interestingly, different from other multi-interest frameworks that cannot beat GRU4Rec, Re4 improves GRU4Rec \wrt NDCG and Hit Rate. On MovieLens with limited items, the larger improvement \wrt NDCG probably indicates that the backward flow helps to learn fine-grained multi-interests (\eg, basketball, and football), which are harder to learn than coarse-grained interests (\eg, sports equipment, and electronic products). With fine-grained multi-interests, Re4 places positive test items in front of the others in the ranking list with confidence, and thus achieving higher NDCG. Nevertheless, Re4 obtain larger performance gains on larger datasets Amazon, demonstrating the practical merits of Re4.
\end{itemize}

\begin{table}[!t]
\centering
    \caption{Analysis of the number of interests $N_z$, which is a hyper-parameter defined in Section \ref{sec:mie}. }
{\setlength{\tabcolsep}{0.7em}\renewcommand{\arraystretch}{1.0}\begin{tabular}{ll cccc}
\toprule

  Model	& Metric  & $N_z=2$ & $N_z=4$ & $N_z=6$  & $N_z=8$   \\
    \midrule
   
  \multirow{6}{*}{Amazon} 
	 & 	R@20	 & 	0.0728	 & 	0.0745	 & 	0.0769 & 	\textbf{0.0771} \\
	 & 	R@50	 & 	0.1033	 & 	0.1105	 & 	\textbf{0.1156}	 & 	0.1155 \\
	 & 	N@20	 & 	0.1239	 & 	0.1277	 & 	0.1298	 & 	\textbf{0.1304} \\
	 & 	N@50	 & 	0.1704	 & 	0.1812	 & 	0.1879 & 	\textbf{0.1883} \\
	 & 	H@20	 &  0.1494	 & 	0.1575	 & 	0.1606 & 	\textbf{0.1627} \\
	 & 	H@50	 & 	0.2063	 & 	0.2220	 & 	0.2311 & 	\textbf{0.2326} \\
    \midrule

    \multirow{6}{*}{MovieLens} 
	 & 	R@20	 & 	0.1007	 & 	0.1121 & 	\textbf{0.1128}	 & 	0.1117 \\
	 & 	R@50	 & 	0.1831	 & 	\textbf{0.2083}	 & 	0.2060	 & 	0.2048 \\
	 & 	N@20	 & 	0.4335	 & 	0.4615	 & 	\textbf{0.4648}	 & 	0.4581 \\
	 & 	N@50	 & 	0.5761	 & 	\textbf{0.6177}	 & 	0.6070 & 	0.6067 \\
	 & 	H@20	 & 	0.7748	 & 	0.7765	 & 	0.7980 & 	\textbf{0.8048} \\
	 & 	H@50	 & 	0.9040	 & 	0.9189	 & 	0.9238	 & 	\textbf{0.9288} \\

    \bottomrule
\end{tabular}}
    \label{tab:numinterest}
\end{table}
%

\begin{table}[!t]
\centering
    \caption{ Ablation studies by progressively adding proposed backward flows to the base model.}
{\setlength{\tabcolsep}{0.65em}\renewcommand{\arraystretch}{1.0}\begin{tabular}{l ccc ccc}
\toprule
Model
   & R@20 & N@20  & H@20  & R@50 & N@50  & H@50  \\
    \midrule
    Base    &   0.128   &   0.274   &   0.384    &   0.207   &  0.402    &   0.529        \\
    \ +$\mathcal{L}_{Att}$    &   0.133   &  0.287    &   0.400    &   0.220   &  0.427    &   0.573        \\
    \ \ +$\mathcal{L}_{CL}$    &   0.135   &   0.296   &  0.412     &   \textbf{0.222}   &  0.433    &   \textbf{0.575}        \\
    \ \ \ +$\mathcal{L}_{CT}$    &   \textbf{0.139}   &   \textbf{0.314}   &   \textbf{0.421}    &   0.220   &  \textbf{0.441}    &   0.570         \\
    \bottomrule
\end{tabular}}%

    \label{tab:ablation}
\end{table}

\begin{table}[!t]
\centering
    \caption{ Analysis of the positive/negative selection threshold $\gamma_c$ in Re-contrast, as defined in Equation \ref{eq:pos}. }
{\setlength{\tabcolsep}{0.65em}\renewcommand{\arraystretch}{1.0}\begin{tabular}{l ccc ccc}
\toprule
$\gamma_c$ &	R@20 &	R@50 &	N@20 &	N@50 &	H@20 &	H@50 \\
    \midrule
1/2 &	0.126 &	0.209 &	0.272 &	0.408 &	0.383 &	0.547 \\
1/4 &	0.128 &	0.221 &	0.277 &	0.425 &	0.387 &	0.567 \\
1/16 &	0.134 &	0.222 &	0.291 &	0.433 &	0.406 &	0.582 \\
1/32 &	0.126 &	0.208 &	0.276 &	0.407 &	0.387 &	0.549 \\
Ada &	0.139 &	0.220 &	0.314 &	0.441 &	0.421 &	0.570 \\
    \bottomrule
\end{tabular}}%

    \label{tab:ablation}
\end{table}

\subsection{In-depth Analysis (RQ2)}

\subsubsection{Analysis of the number of user interest embeddings.} \label{sec:numinterest} To have a comprehensive analysis of how the number of user interest embeddings affects the performance of Re4, we conduct experiments on a large-scale dataset Amazon with diverse items (\ie, 313, 966) and a small-scale dataset MovieLens with limited items (\ie, 3, 900). According to the results shown in Table \ref{tab:numinterest}, we have several findings:
\begin{itemize}[leftmargin=*,labelsep=0.8mm]
\renewcommand\labelitemi{{\boldmath$\cdot$}}
	\item Increasing the amount of interest embeddings mostly leads to performance gains (\eg, $2 \rightarrow 4 \rightarrow 6 \rightarrow 8$ on Amazon and $2 \rightarrow 4,6,8$ on MovieLens). When increasing the amount of interest embeddings, a multi-interest framework probably yields trivial multi-interest embeddings that are similar to each other, or incorrect interest embeddings that are semantically irrelevant to their corresponding representative items. Such ways of modeling hinder performance gains. As such, the performance gain basically demonstrates that Re4 can help to learn effective multi-interest embeddings through the backward flow, \ie, Re-contrast, Re-construct, and Re-attend.
	\item When increasing the number of interests, Re4 consistently improves the performance on the Amazon dataset while there are minor performance drops \wrt several metrics on the MovieLens dataset. This result generally indicates the practical merits of Re4 for large-scale recommender systems where the information overload problem is more severe. This finding is consistent with the results shown in Table \ref{tab:comparison} and analyzed in Section \ref{sec:overall}.
	\item Although there is a performance drop on small-scale dataset MovieLens when the number of interest increases, \ie, $4 \rightarrow 6, 6 \rightarrow 8$, the change is relatively lower than the performance gain obtained in $2 \rightarrow 4$. Jointly analyzing the results of larger-scale dataset Amazon, the performance of Re4 is generally less sensitive to over-increasing the hyper-parameter, \ie, the number of interests. This observation further demonstrates the practical merits of Re4.
\end{itemize}

\begin{table}[!t]
\centering
    \caption{ Analysis of how loss coefficients $\lambda_{Att}$, $\lambda_{Att}$, and $\lambda_{Att}$ affect the performance of Re4. For each experiment, we add one loss function to the base model and alter the corresponding coefficient. }
{\setlength{\tabcolsep}{1em}\renewcommand{\arraystretch}{1.0}\begin{tabular}{ll ccc}
\toprule

  Model	&  Coeff.  & H@20 & R@20  & N@20   \\
  \midrule
      Base  &   NA   &   0.3838   &   0.1277    &   0.2736    \\

    \midrule
   
    \multirow{3}{*}{w. $\mathcal{L}_{Att}$} 
	 & 	1	 & 	0.3868	 & 	0.1287	 & 	0.2759 \\
	 & 	0.1	 & 	0.4042	 & 	0.1302	 & 	0.2843 \\
	 & 	0.01	 & 	0.4002	 & 	0.1334	 & 	0.2875 \\
    \midrule
    \multirow{3}{*}{w. $\mathcal{L}_{CL}$} 
	 & 	1	 & 	0.4039	 & 	0.1328	 & 	0.2846 \\
	 & 	0.1	 & 	0.4049	 & 	0.1335	 & 	0.2874 \\
	 & 	0.01	 & 	0.4056	 & 	0.1351	 & 	0.2905 \\
    \midrule	 
    \multirow{3}{*}{w. $\mathcal{L}_{CT}$} 
	 & 	1	 & 	0.4042	 & 	0.1341	 & 	0.2861 \\
	 & 	0.1	 & 	0.4066	 & 	0.1325	 & 	0.2883 \\
	 & 	0.01	 & 	0.3995	 & 	0.1331	 & 	0.2862 \\

    \bottomrule
\end{tabular}}
    \label{tab:coeff}
\end{table}
%

\subsubsection{Ablation Studies.} To investigate how different backward flows (\ie, Re-contrast, Re-construct, and Re-attend) affect the performance of Re4, we conduct ablation studies by progressively adding three strategies to the base model. The results on the Gowalla dataset are shown in Table \ref{tab:ablation}.
\begin{itemize}[leftmargin=*,labelsep=0.8mm]
\renewcommand\labelitemi{{\boldmath$\cdot$}}
	\item \textbf{$\mathcal{L}_{Att}$} denotes the loss function of the Re-attend backward flow, which explicitly guarantees that the dot product similarities between interests and items are consistent with the attention weights in the forward flow. The performance gain demonstrates the importance of such consistency. We attribute the improvement to that many matching models, including the base model ComiRec-SA, make recommendations based on the dot product similarities between interest embeddings and item embeddings. 
	\item \textbf{$\mathcal{L}_{CL}$} refers to the Re-contrast backward flow, which drives interest embeddings to be distinct from each other. Adding $\mathcal{L}_{CL}$ leads to consistent recommendation performance improvement, validating the necessity of the Re-contrast backward flow for the multi-interest framework.
	\item \textbf{$\mathcal{L}_{CT}$} is designed to ensure that each interest embedding can semantically reflect the corresponding representative items. Adding $\mathcal{L}_{CT}$ leads to more gains \wrt NDCG. The reason might be that semantic reflection can help the interest embeddings to have a fine-grained understanding of the representative items rather than the coarse-grained understanding such as the correlation. Fine-grained understanding is further graceful for distinguishing candidates at the top ranks, and gives accurate items higher ranks. Another evidence for the analysis is that the improvement on Top 20 candidates are higher than Top 50 candidates. This merit is essential since, in many real-world recommender systems, users generally give fewer chances for items that are left behind.
\end{itemize}

%
%
%
%
%

\begin{figure}[!t] \begin{center}
\begin{subfigure}{.23\textwidth}
	\includegraphics[width=1\linewidth]{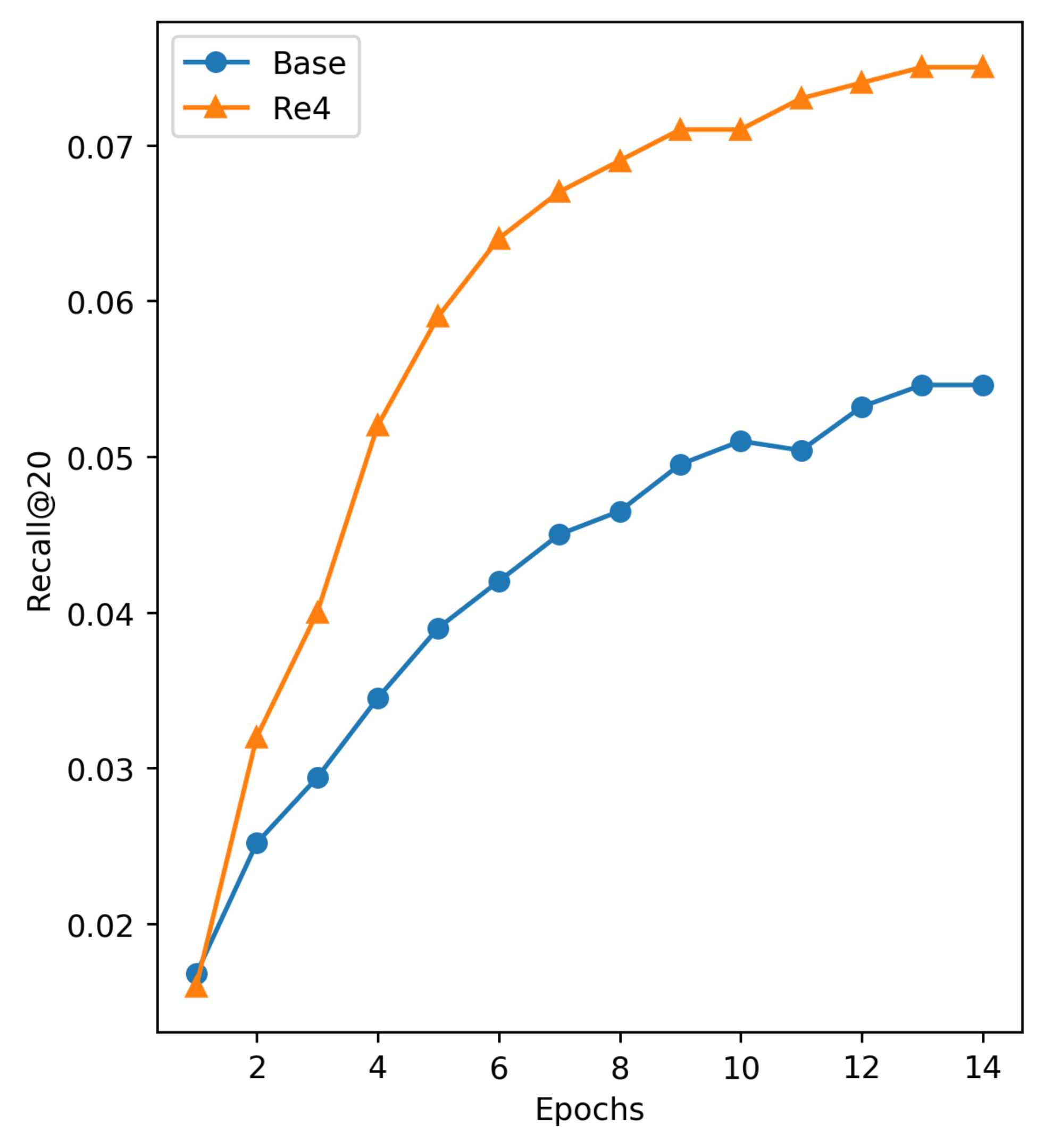}
    \caption{
    \footnotesize{Amazon}
    	}
\label{fig:tsnebase}
\end{subfigure}
\begin{subfigure}{.23\textwidth}
	\includegraphics[width=1\linewidth]{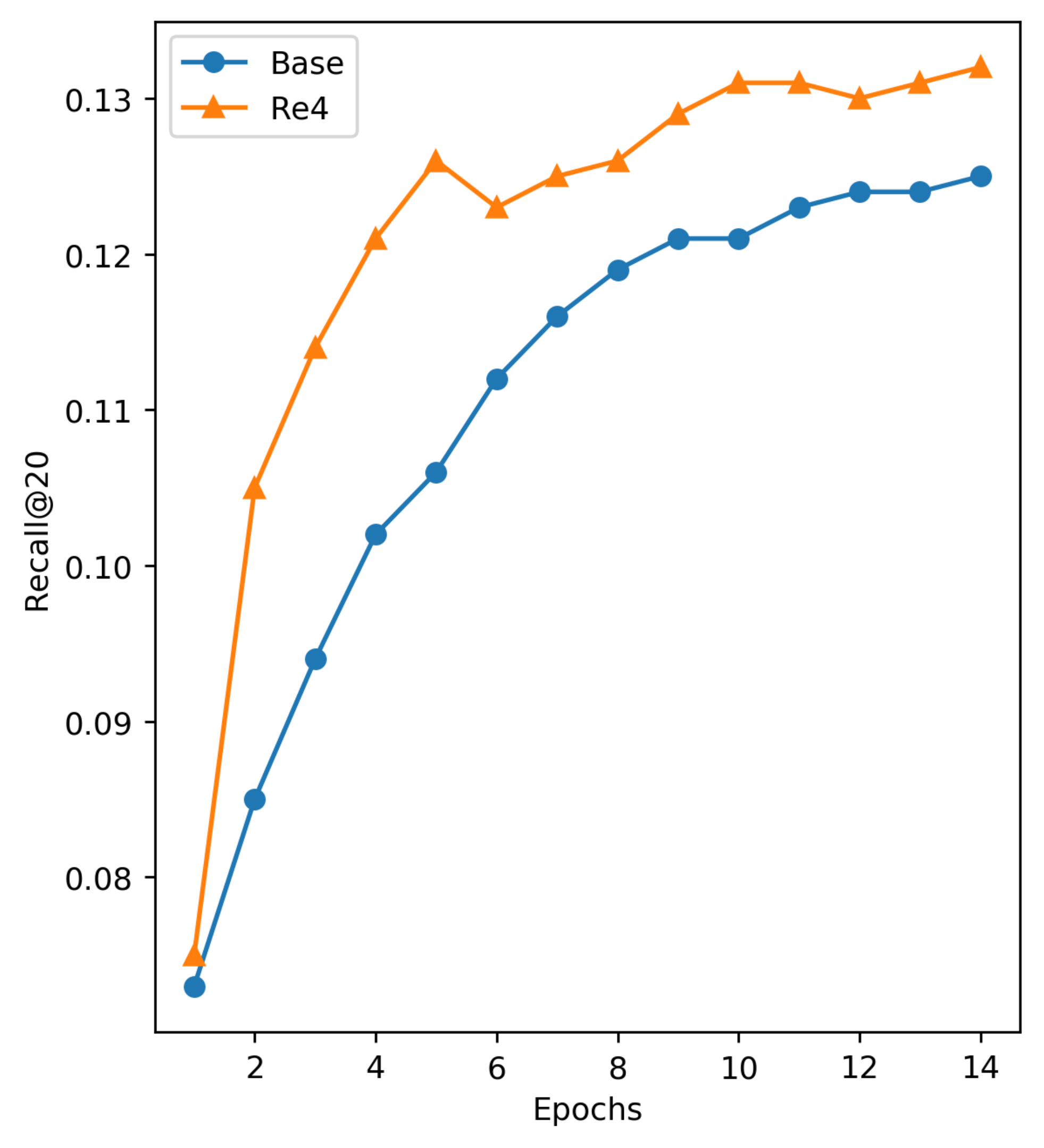}
    \caption{
   \footnotesize{Gowalla}
    	}
\label{fig:tsneRe4}
\end{subfigure}
    \caption{
    Test performance across different epochs of Base without backward flow and Re4.
    	}
    \label{fig:testperepoch}
\end{center} \end{figure}

\subsubsection{Analysis of Loss Coefficients.} Since Re4 introduces new loss functions, we take an analysis of how loss coefficients affect the performance of the corresponding backward flow. Specifically, for each experiment, we solely add one loss function among $\mathcal{L}_{Att}, \mathcal{L}_{CL}, \mathcal{L}_{CT}$ to the base model, and manually set its coefficient to $1, 0.1, 0.001$. The results on Gowalla dataset are shown in Table \ref{tab:coeff}. When altering the loss coefficients, there are minor performance changes in most cases. In other words, the recommendation performance is insensitive to the hyper-parameters. An exception is the $\mathcal{L}_{Att}$ loss function, which leads to less effective performance when the corresponding coefficient is high $\lambda_{Att}=1$. The reason might be that over-regularization on the attention weights probably limits the expressive power of attention mechanism in the forward flow. Still, all variants of Re4 perform better than the base model, demonstrating the merits of three backward flows.

\subsubsection{Analysis of $\gamma_c$ in Re-contrast.} In positive/negative items selection of Re-contrast, we introduce a hyperparameter $\gamma_c$ as the threshold, as defined in Equation \ref{eq:pos}. In practice, we choose an adaptive $\gamma_c=1/N_x$ to somehow balance the number of positives and negatives. To have an analysis of the how $\gamma_c$ affects the performance of Re4, we conduct experiments with $\gamma_c=\{ 1/2, 1/4, 1/16, 1/32 \}$. Note that large $\gamma_c = 1/2$ and small $\gamma_c = 1/32$ correspond skewed cases with significantly more negatives and positives, respectively. $\gamma_c = 1/16$ yields a relatively proper ratio of positives to negatives, leading to better performance than the skewed cases. The adopted adaptive strategy $\gamma_c=1/N_x$ achieves better performance than the fixed value $\gamma_c = 1/16$ in most cases.

\subsubsection{Test Performance across Epochs.} As shown in Figure \ref{fig:testperepoch}, the test performance for each epoch is evaluated in terms of Recall@20.  According to the results, Re4 not only achieves better performance when convergence but also obtains more substantial improvement in the early stage of training.
This observation demonstrates that the backward flows in Re4 are effective in learning multi-interest representations. This merit is critical in real-world recommender systems where there are numerous interactions logged daily, and models are more likely to be trained with fewer epochs.

\subsection{Effect on Representation (RQ3)} 

\vpara{Case Study.} We are interested in how the proposed backward flows facilitate multi-interest representation learning in the embedding space. As such, we train the two-interest version of both the base model (ComiRec-SA) and Re4. We randomly sample six users and their corresponding items, and obtain users' multi-interest embeddings and item embeddings. We perform t-SNE transformation onto these embeddings and plot the results in Figure \ref{fig:tsnebase} and Figure \ref{fig:tsneRe4} for Base and Re4, respectively. Both users and items are randomly sampled from the test set and are unseen during training, which helps to better reveal the generalization ability. We have the following findings:
\begin{itemize}[leftmargin=*,labelsep=0.8mm]
\renewcommand\labelitemi{{\boldmath$\cdot$}}
	\item Overall, users with their corresponding items exhibit more noticeable clusters in Re4. The base model is more likely to yield trivial multi-interest embeddings. For instance, the two interest embeddings of user 29742 present no significant difference \wrt the distance to their corresponding items. As for user 28514, while one interest embedding (top) is close to its corresponding items, another interest embedding (bottom) is far away from the items. These inferior results probably verify the analysis in Section \ref{sec:numinterest}. As a remedy, the proposed Re-contrast backward flow can drive multiple interest embedding to be discrepant while the Re-attend and Re-construct backward flows can ensure interest embeddings are closed to their representative items both in correlation and in semantics.
	\item For Re4, two interest embeddings with their closest items mostly exhibit two fine-grained clusters. For example, for user 28027, 28286, and 28233, the two interests' representations are not only distinct from each other in the embedding space, but also with exclusive test items around. Meanwhile, each user's multiple interests exhibit a larger distance with other users' items and interests than the distance with items and interests of the same user. These results jointly demonstrate that Re4 learns effective multiple embeddings that can represent different aspects of interests.
\end{itemize}

\vpara{Quantitative Result.} Besides the above qualitative results, we also provide quantitative results as follows:
1) We perform clustering on the representations of all users' interests and items. We evaluate the ratio of positive items being in the cluster of their corresponding interest, \ie, \textbf{INTER} (-User).
2) For each user, we perform clustering on the representation of his/her interest and items representations. We evaluate the ratio of his/her interests being in different clusters, \ie, \textbf{INTRA} (-User). 
We use multiple cluster initialization (CI) methods and multiple clustering methods (CM). We use the negative of recsys’s similarity function (dot product) as the distance metric. According to Table \ref{tab:representation}, we have similar observations as in the case study: 
1) Users and their corresponding items exhibit more noticeable clusters in Re4 (with high INTER); and 2) different interest embeddings with their closest items are more likely to exhibit fine-grained clusters in Re4 (with high INTRA).

\begin{figure}[!t] \begin{center}
\begin{subfigure}{.43\textwidth}
	\includegraphics[width=0.95\linewidth]{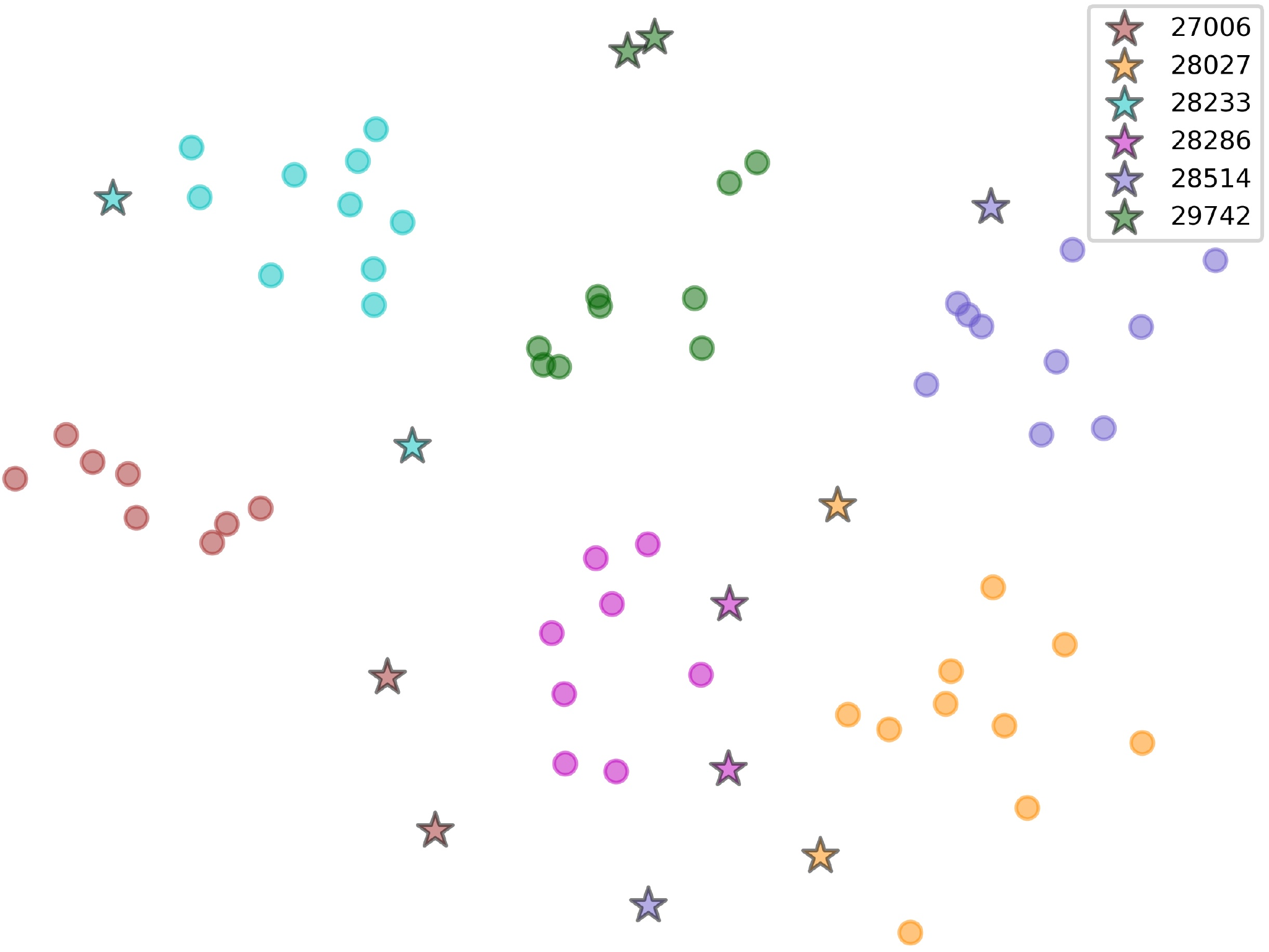}
    \caption{
    \footnotesize{Base}
    	}
\label{fig:tsnebase}
\end{subfigure}
\begin{subfigure}{.43\textwidth}
	\includegraphics[width=0.95\linewidth]{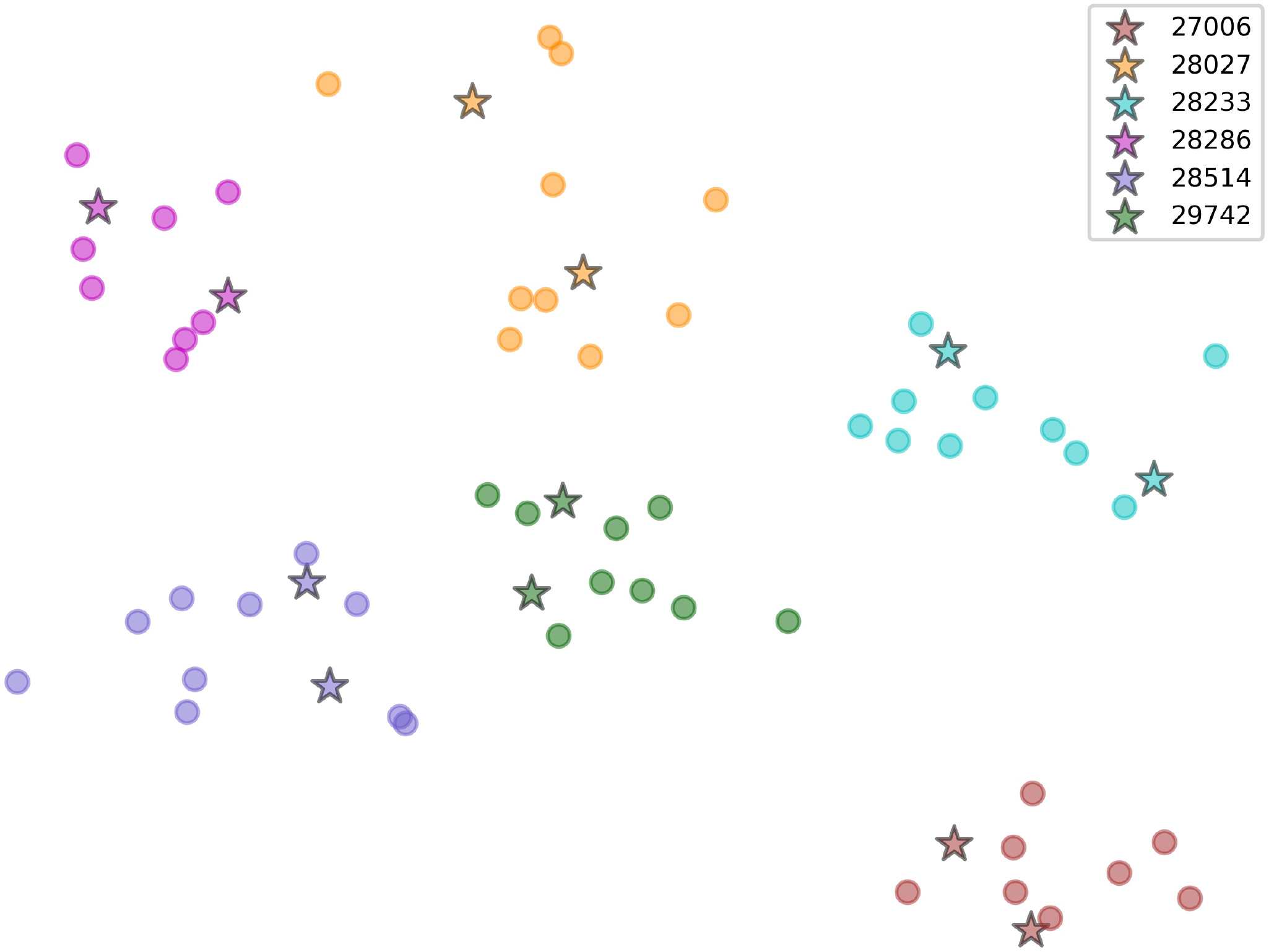}
    \caption{
   \footnotesize{Re4}
    	}
\label{fig:tsneRe4}
\end{subfigure}
    \caption{
    The visualization displays the multi-interests ($\star$) of some randomly sampled test users, and some corresponding items ($\bullet$ of the same color). We perform t-SNE transformation on the multi-interest embeddings and item embeddings learned by the base model without backward flow and Re4.
    	}
    \label{fig:tsne}
\vspace{-0.2cm}
\end{center} \end{figure}

\begin{table}[!t]
\centering
    \caption{ Quantitative analysis of representations learned by the Base model and Re4.}
{\setlength{\tabcolsep}{0.7em}\renewcommand{\arraystretch}{1.0}\begin{tabular}{ll  cc|cc}
\toprule

   	& CI  & \multicolumn{2}{c}{K-means++} & \multicolumn{2}{c}{User Interests}   \\
    CM	& Metric  & Base & Re4 &  Base & Re4   \\
    \midrule
    
    \multirow{2}{*}{K-means} 
	 & 	INTER	 & 	20.32	 & 	23.03	 & 	26.91	 & 	32.98 \\
	 & 	INTRA	 & 	35.10	 & 	37.14	 & 	35.80	 & 	38.65 \\
	     \midrule
    \multirow{2}{*}{FCM} 
	 & 	INTER	 & 	84.50	 & 	88.89	 & 	74.58	 & 	84.72 \\
	 & 	INTRA	 & 	36.47	 & 	38.14	 & 	37.84	 & 	39.48 \\	
	     \midrule
    \multirow{2}{*}{X-means} 
	 & 	INTER	 & 	26.29	 & 	31.22	 & 	29.18	 & 	35.15 \\
	 & 	INTRA	 & 	36.07	 & 	37.78	 & 	37.47	 & 	39.05 \\

    \bottomrule
\end{tabular}}
\vspace{-0.2cm}
    \label{tab:representation}
\end{table}
%

\section{Conclusion and Future Work} \label{sec:conclusion}

In this paper, we investigate how we can model and leverage the backward flow (interests-to-items) for multi-interest recommendation. We devise the Re4 framework that incorporates three backward flows, \ie, Re-contrast, Re-construct, and Re-attend. In essence, Re4 facilitates the multi-interest representation learning to 1) capture diverse aspects of interest; 2) semantically reflect the corresponding representative items; and 3) make the attention weights in the forward flow consistent with interest-item correlation \wrt the final recommendation. We conduct extensive experiments on three real-world datasets, providing insightful analyses on the rationality and effectiveness of Re4. This work was an initiative to construct the forward-backward paradigm for multi-interest recommendation. Remarkably, the backward flow does not affect inference, which is essential for industrial recommender systems. We believe that the novel paradigm can be inspirational to future developments. Obviously, there is more to explore on backward flow strategies. For example, we can extend Re4 to content-based recommenders and explore strategies for various auxiliary features.

\section{ACKNOWLEDGMENTS}

\begin{sloppypar}
The work is supported by the National Natural Science Foundation of China (No. 62037001,61836002,62072397), National Key R\&D Program of China (No. 2020YFC0832500), Zhejiang Natural Science Foundation (No. LR19F020006), and Project by Shanghai AI Laboratory (No. P22KS00111).
\end{sloppypar}

\balance

\bibliographystyle{ACM-Reference-Format}
\bibliography{sections/9.citations}


\end{document}